\documentclass[pra,twocolumn,showpacs,superscriptaddress]{revtex4-1}
\usepackage{graphicx}
\usepackage{bm}
\usepackage{amssymb}
\usepackage{color}
\usepackage{amsmath}
\usepackage{amstext}
\usepackage{latexsym}
\usepackage[usenames,dvipsnames]{xcolor}
\usepackage[colorlinks=true,citecolor=purple,linkcolor=RubineRed,urlcolor=purple]{hyperref}
\usepackage{bbold}
\usepackage{float}
\usepackage{enumerate}
\usepackage{braket}

\definecolor{pink}{rgb}{1,0.078,0.57}

\definecolor{green}{rgb}{0,0.7,0.9}

\def\tr{{\rm Tr}}

\newcommand{\beq}{\begin{equation}}
\newcommand{\eeq}{\end{equation}}
\newcommand{\beqa}{\begin{eqnarray}}
\newcommand{\eeqa}{\end{eqnarray}}

\newcommand{\srho}{\tilde{\rho}}

\newcommand{\avs}[1]{\mathbb{E}(#1)} 

\newcommand{\be}[1]{\mathbf{#1}}
\newcommand{\mc}[1]{\mathcal{#1}}
\newcommand{\mr}[1]{\mathrm{#1}}
\newcommand{\ts}[1]{\textsc{#1}}

\newcommand{\ha}[1]{\hat{#1}}
\newcommand{\ti}[1]{\tilde{#1}}

\newcommand{\mbb}[1]{\mathbb{#1}}
\newcommand{\bs}[1]{\boldsymbol{#1}}

\newcommand{\dg}{^{\dagger}}
\newcommand{\Tr}{\mathrm{Tr}}

\newcommand{\uvrho}{\tilde{\varrho}}
\newcommand{\urho}{\tilde{\rho}}

\newcommand{\uL}{\tilde{\mathcal{L}}}

\newcommand{\aritra}{\color{black}}

\begin{document}

\title{Quantum Dynamics with Stochastic Non-Hermitian Hamiltonians}

\author{Pablo Martinez-Azcona\href{https://orcid.org/0000-0002-9553-2610}{\includegraphics[scale=0.05]{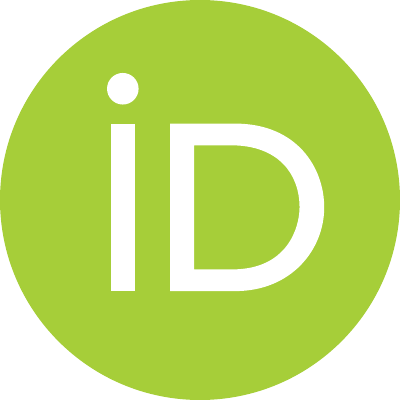}}}
\affiliation{Department of Physics and Materials Science, University of Luxembourg, L-1511 Luxembourg}
\author{Aritra Kundu\href{https://orcid.org/0000-0001-7476-8811}{\includegraphics[scale=0.05]{orcidid.pdf}}}
\affiliation{Department of Physics and Materials Science, University of Luxembourg, L-1511 Luxembourg}
\author{\\Avadh Saxena\href{https://orcid.org/0000-0002-3374-3236}{\includegraphics[scale=0.05]{orcidid.pdf}}}
\affiliation{Theoretical Division,
Los Alamos National Laboratory, Los Alamos, New Mexico 87545, USA}
\author{Adolfo del Campo\href{https://orcid.org/0000-0003-2219-2851}{\includegraphics[scale=0.05]{orcidid.pdf}}}
\affiliation{Department of Physics and Materials Science, University of Luxembourg, L-1511 Luxembourg}
\affiliation{Donostia International Physics Center,  E-20018 San Sebasti\'an, Spain}
\author{Aur\'elia Chenu\href{https://orcid.org/0000-0002-4461-8289}{\includegraphics[scale=0.05]{orcidid.pdf}}}
\affiliation{Department of Physics and Materials Science, University of Luxembourg, L-1511 Luxembourg}


\begin{abstract}

We study the quantum dynamics generated by a non-Hermitian Hamiltonian subject to stochastic perturbations in its anti-Hermitian part, describing fluctuating gains and losses.  
{The dynamics averaged over the noise is described by an `anti-dephasing' master equation.} 
We characterize the resulting state evolution and analyze its purity. The properties of such dynamics are illustrated in a stochastic dissipative qubit. 
{Our analytical results show that adding noise allows for a rich control of the dynamics, stabilizing the lossy state and making state purification possible to a greater 
variety of steady states.}

\end{abstract}

 \maketitle
 \textit{Introduction---}
Quantum systems are always in contact with an environment. For this reason, any realistic description of quantum dynamics should include the effects of such coupling. The theory of open quantum systems (OQS) offers a wide range of approaches to this end \cite{BreuerBook, RivasBook}. Two prominent approaches involve non-Hermitian (NH) \cite{ashida_non-hermitian_2020, PK98, Moiseyev} and stochastic \cite{budini_quantum_2001, chenu_quantum_2017, kiely_exact_2021} Hamiltonians.
NH Hamiltonians are generally associated with effective descriptions that account only for a subset of the total number of degrees of freedom of a quantum mechanical system. They can be obtained by unraveling the Gorini-Kossakowski-Sudarshan-Lindblad (GKSL)  equation \cite{Gorini76, Lindblad76} and post-selecting the trajectories with no quantum jumps \cite{ashida_non-hermitian_2020, roccati_non-hermitian_2022}, or using projector techniques, as often done in quantum optics, nuclear physics, and molecular quantum chemistry \cite{PK98,Moiseyev, Feshbach1958, Feshbach1962, CohenTannoudji1998}.   
Recently, non-Hermitian Hamiltonians have regained interest because of the novel 
physical phenomena and topological phases they display \cite{ShenZhenFu18,Zongping18,Kawabata18,YaoWang18,Takata18,Yang18, guo_NHBoost_23}, and of their relation to the quantum Zeno effect \cite{ALLCOCK1969253, Schulman98,Streed06,Echanobe08,Dubey_Zeno_2023,Snizhko_Zeno_2020}.

NH Hamiltonians generally exhibit complex energy eigenvalues in which the imaginary part describes the lifetime of a state \cite{Gamow1928, CohenTannoudji1998}. Time evolution ceases to be trace-preserving due to the leakage of probability density between the subspace under consideration and that associated with the remaining degrees of freedom. 
The evolution is made trace-preserving (TP) by renormalizing the density matrix at all times, which renders the evolution nonlinear in the state
\cite{carmichael_statistical_2008,BrodyGraefe12, Alipour2020,geller_fast_22, Rembielinski2021nonlinearextension}. TP nonlinear dynamics has been realized experimentally by post-selection of 
dissipative evolution, i.e., discarding the trajectories with quantum jumps \cite{naghiloo_quantum_2019, chen_quantum_2021}. In this setting, the renormalization of the trace comes naturally as a restriction to the subset of trajectories with no quantum jumps. 
This allowed studying a plethora of dissipation-induced phenomena \cite{abbasi_topological_NHQubit, erdamar_constraining_2024, Harrington2022, chen_decoherence_2022, wang_dissipative_2023}. 

Stochastic Hamiltonians allow for a treatment of OQS in which the effect of the environment is 
{modeled} through the noise statistics 
\cite{VanKampen2011,budini_quantum_2001,kiely_exact_2021}. These appear naturally in the stochastic Schrödinger equations obtained by unraveling the GKSL equation in quantum trajectories \cite{wiseman_quantum_1996, wiseman_quantum_2009, carmichael_statistical_2008}. Stochastic Hamiltonians have been used to engineer many-body and long-range interactions \cite{chenu_quantum_2017}, and their behavior beyond the noise average relates to quantum information scrambling \cite{martinez-azcona_stochastic_2023}. 

While the stability of certain features of NH Hamiltonians under noise has recently been investigated \cite{wiersig_robustness_2020}, non-Hermitian Hamiltonians are typically studied in a deterministic setting. In this Letter, we go beyond this paradigm by studying the dynamics of NH Hamiltonians subject to classical noise. This is particularly relevant since current experimental realizations of NH dynamics use Noisy Intermediate Scale Quantum (NISQ) devices \cite{Preskill2018quantumcomputingin}. We choose to focus on the effect of fluctuations on the anti-Hermitian part, which 
arise from noise in experimental setups \cite{naghiloo_quantum_2019} 
or in homodyne detection of the degenerate parametric oscillator \cite{carmichael_open_1993}, the possible experimental setups are detailed in App. \ref{app:exp_real}.
By considering the stochastic dynamics of a fluctuating non-Hermitian system, we unveil a new kind of {averaged} master equation, that we call `anti-dephasing' because of the structure of the dissipator---which involves anticommutators instead of the commutators commonly encountered in the GKSL dephasing master equation.
%
We derive an exact {evolution equation} 
for the purity and characterize its dynamics. Specifically, we identify the stable steady states in terms of the eigenvalues and eigenstates of the Liouvillian. 
We illustrate the anti-dephasing dynamics in an experimentally-feasible stochastic dissipative qubit \cite{naghiloo_quantum_2019, quinn2023observingsuperquantumcorrelationsexceptional} (cf. App. \ref{app:exp_real}),  in which case we find a novel noise-induced phase, where the noise stabilizes the originally lossy state. 

\emph{Dynamics under anti-Hermitian fluctuations---}
We consider a system with Hermitian Hamiltonian $\ha H_0$ that is subject to classical noise coupled to the anti-Hermitian operator $i\ha L$, i.e.,
\begin{equation} \label{eq:stoch_H}
    \ha H_t = \ha H_0 - i (1+\sqrt{2 \gamma}  \xi_t) \ha L.
\end{equation}
This describes fluctuations around the mean in the Hermitian $\ha L =\hat{L}^\dagger$, positive $\ha L\geq 0$ operator, with standard deviation $\sqrt{2 \gamma}$, and could be amenable to experimental implementation (cf. App. \ref{app:exp_real}).
The classical noise is taken as Gaussian real white noise for simplicity, characterized by $\avs{\xi_t  }=0$ and $\avs{\xi_t  \xi_{t'}}= \delta(t-t')$, where $\avs{\bullet}$ denotes the classical stochastic average. 
The (unnormalized) system density matrix for a single trajectory $\uvrho_t$ evolves, over a small time increment, as $\uvrho_{t+ \mr dt}=\ha U_{\mr dt}\uvrho_t \ha U\dg_{\mr dt}$, where the propagator 
  $\ha U_{\mr dt} = \exp(- i \ha H_0 \mr dt - \ha L \mr dt-\sqrt{2 \gamma} \ha L \mr d W_t )$ 
depends on  the Wiener process $\mr d W_t  = \xi_t  \mr dt$. For Gaussian white noise, the latter obeys It\=o's rules \cite{gardiner_handbook_1985}: $\mr d W_t^2 = \mr dt$ and $(\mr d W_t )^n=0, \; \forall \;  n >2$. Expanding the exponential dictating the evolution of the density matrix with these rules, we find a stochastic differential equation (SDE) for its evolution,
\begin{align} \notag 
    \mr d \uvrho_t =  &\left( - i [\ha H_0, \uvrho_t] -  \{\ha L, \uvrho_t\}+\gamma \big\{\ha L, \{\ha L, \uvrho_t\}\big\} \right)\mr d t  \\ &- \sqrt{2 \gamma} \{\ha L, \uvrho_t\} \mr d W_t.
\end{align}
We denote  $\urho \equiv \avs{\uvrho}$  the noise-averaged density matrix and use the tilde $\tilde{\bullet}$ to highlight that it is not normalized.
Trace preservation can be imposed either by (i)  renormalizing all single-trajectories 
and then averaging over the noise  \cite{carmichael_statistical_2008, wiseman_quantum_2009, Dubey_Zeno_2023,benoist_emergence_2021}, or (ii) first averaging over the noise, and only then imposing trace preservation. 
{ These two approaches will generally yield different dynamics. 
We focus on the latter to capture the evolution of ensembles of trajectories analytically and defer the former for future analysis.}

The average over the noise of the unnormalized evolution reads 
\begin{align} \label{eq:durhodt}
    \frac{\mr d\urho_t}{\mr dt} = - i [\ha H_0, \urho_t] - \{\ha L, \urho_t\}+ \gamma\big \{\ha L, \{\ha L, \urho_t\}\big\} \equiv \uL[\urho_t],
\end{align}
for the noise-averaged density matrix, where $\uL[\bullet]$ denotes the 
Liouvillian superoperator, which is not necessarily trace-preserving. This master equation can be formally solved as $\urho_t = e^{\uL t}[\ha \rho_0]$. 
Imposing trace preservation at the average level yields the nonlinear master equation 
\begin{align} \label{BNGL_master_eq}
    \frac{\mr d \hat{\rho}_t}{\mr dt} = &\; \uL[ \hat{\rho}_t]-\Tr(\uL[ \hat{\rho}_t]) \hat{\rho}_t\\  \notag
    = &- i [\ha H_0,  \hat{\rho}_t] -\{\ha L, \hat{\rho}_t\}+ \gamma \big \{\ha L, \{\ha L,  \hat{\rho}_t\}\big \}\\ \notag &+2 \Tr(\ha L  \hat{\rho}_t) \hat{\rho}_t- 4 \gamma \Tr(\ha L^2  \hat{\rho}_t)  \hat{\rho}_t,
\end{align}
the solution of which is given by normalizing the non-trace-preserving (nTP) evolution, i.e., 
    $ \hat{\rho}_t \equiv \frac{\urho_t}{\Tr(\urho_t)}=e^{\uL t}[\hat{\rho}_0]/ \Tr(e^{\uL t}[\hat{\rho}_0]).$
Since the NH Hamiltonian is stochastic, the evolution leads to quantum jumps---terms of the form $\gamma \ha L \ha \rho_t \ha L$. 
Note that this master equation is not of GKSL form; instead,  the jump operators act on the density matrix through a double anticommutator and do not conserve the norm of the state. We term this new dissipator `anti-dephasing'. A more standard form is detailed in App.~\ref{app:standard_form}. 
Similar equations, with a double anticommutator but with a minus sign, appear when studying the adjoint Liouvillian with fermionic jumps and evolving operators \cite{Liu:2022god, Clark_exact_2010}.

Note that, in general, the unnormalized evolution \eqref{eq:durhodt} can be trace-decreasing (TD), trace-preserving (TP), or trace-increasing (TI). While TD dynamics can easily be interpreted through post-selection, e.g., with \textit{hybrid Liouvillians} \cite{minganti_hybrid_20, minganti_EP_19, gupta_jumps_24}, TI evolutions are conceptually challenging; even if sometimes deemed non-physical \cite{nielsen_quantum_2010} they can describe a biased ensemble of trajectories \cite{garrahan_thermodynamics_2010, Lebowitz1999, esposito_noneq_09, carollo_rare_18, Carollo2021, cech_thermodynamics_2023}. 
Remarkably, the present TI dynamics can always be made TD or TP by a gauge transformation in the master equation, namely, using an imaginary offset $-i a$ in \eqref{eq:stoch_H} (cf. App. \ref{app:gauge}). { This is because such a transformation rigidly shifts the Liouvillian spectrum, conserving the relative differences between eigenvalues and, hence, the dynamical features.}

We now characterize the dynamics generated by \eqref{BNGL_master_eq} in terms of the purity evolution and the stable steady states in the general setup before illustrating them in a stochastic dissipative qubit.

\emph{Purity dynamics---} 
The purity $P_t = \Tr(\ha \rho_t^2)$ { 
quantifies how mixed a state is, even when its trace evolves in time. 
For anti-dephasing dynamics, it }evolves according to the differential equation
    \begin{align} \notag
    \partial_t P_t= &-4 \Tr(\ha L \hat{\rho}_t^2) + 4 \gamma \left( \Tr(\ha L^2 \hat{\rho}_t^2) + \Tr(\ha L \hat{\rho}_t \ha L \hat{\rho}_t)\right)\\ \label{eq:tauP}
    &+4 \Tr(\ha L \ha \rho_t)P_t - 8 \gamma \Tr(\ha L^2\ha \rho_t)P_t.
\end{align}
Evaluating this expression at $t=0$ yields the decoherence time $1/\tau_P\equiv \partial_t P_t \vert_{t=0} $.
Whenever the state is pure, $\hat{\rho}_t= \ket{\Psi_t}\bra{\Psi_t}$, the terms coming from the deterministic anti-Hermitian part cancel out
and the purity evolution $\partial_t P_t=-4 \gamma {\rm Var}_{\Psi}(\ha L)$ is  
 dictated by the variance ${\rm Var}_{\Psi}(\ha L)=\braket{\Psi_t|\ha L^2|\Psi_t}-\braket{\Psi_t|\ha L|\Psi_t}^2$.
Thus, initial pure states exhibit a purity decay unless they are eigenstates of $\ha L$, in this case, the purity remains constant at first order in $t$ ($\tau_P\rightarrow\infty$). This result is identical to the decoherence time of pure states in a dephasing channel and quantum Brownian motion \cite{Zurek03, BreuerBook,Beau17decay}. Interestingly, this expression contains out-of-time-order terms reminiscent of OTOC's \cite{martinez-azcona_stochastic_2023} and generalizes the known evolution for NH Hamiltonians \cite{BrodyGraefe12}. 

\emph{Long time dynamics: Stable steady states---} 
To characterize the long-time dynamics, we use the right and left eigendecompositions of the 
Liouvillian superoperator, $\ti{\mc L}[\urho_\nu]= \lambda_\nu \urho_\nu$ and $\uL\dg[\urho_\nu^\ts{(l)}]=\lambda_\nu^* \urho_{\nu}^\ts{(l)}$, respectively. We define the eigenoperator basis as $\ha \sigma_\nu =  \urho_\nu /\Tr( \urho_\nu)$ if the operator has nonzero trace and as $\ha \sigma_\nu =  \urho_\nu$ if the operator is traceless. Furthermore, the eigenstate is \textit{physical} when it has unit trace, $\Tr(\ha \sigma_\nu) =1$, and is positive semidefinite $\ha \sigma_\nu \geq 0$. The basis is ordered such that the physical states appear first $0 \leq \nu < N_\ts{p}$, the traceful states appear second $N_\ts{p}\leq \nu < N_\ts{t}$, and the traceless states come last. In each subset, the states are further ordered by decreasing $\mr{Re}(\lambda_\nu)$.
Assuming the Liouvillian $\uL$ to be diagonalizable, the evolution of the density matrix in this operator basis follows as 
\begin{align}\label{eq4}
    \!\!\!\hat{\rho}_t &= \frac{\sum_{\nu=0}^{N^2-1} c_\nu e^{\lambda_\nu t}  \ha \sigma_\nu}{\sum_{\mu=0}^{N_\ts{t}-1} c_\mu e^{\lambda_\mu t} }\stackrel{t \rightarrow \infty }{\sim} \ha \sigma_{0} {+} e^{-\Delta t} \sum_{\nu \in \mbb M_{2} } \frac{c_\nu e^{i \omega_\nu t}}{c_{0}} \ha \sigma_{\nu} , 
\end{align}
where the coefficients are  $c_\nu = \Tr(\ha \sigma_\nu^\ts{(l)}\hat{\rho}_0 )/\tr(\ha \sigma_\nu^\ts{(l)} \ha \sigma_\nu)$. All the physical states ($\ha \sigma_\nu$ with $\nu < N_\ts{p}$) are steady states. However, only those whose eigenvalues have the largest real part $\ha \sigma_0$ are stable under all perturbations \footnote{The stability and restrictions imposed by complete positivity will be studied in a follow-up article (in preparation).}. 

At long times, the stable steady state $\ha \sigma_{0}$ is the eigenoperator associated with the eigenvalue with the largest real part over which there is initial support, i.e., $\mr{Re}(\lambda_{0})= \max_\nu \mr{Re}(\lambda_{\nu})$ such that $c_{\nu}\neq 0 $. The corrections to this state are suppressed at a rate dictated by the \textit{dissipative gap} $\Delta = \mr{Re}(\lambda_0-\lambda_1)$. They can oscillate with a frequency $\omega_\nu = \mr{Im}(\lambda_\nu)$, depending on the presence or absence of complex eigenvalues in the set of states with the second largest real part, denoted by $\mbb M_2$.

The long-time limit above \eqref{eq4} assumes that the largest eigenvalue is real and non-degenerate. In a more general case, denoting by $\mbb M_1$ the set of states with the largest real part among those on which the initial state has nonzero support, the steady state follows as $\hat{\rho}^\ts{s} = \frac{\sum_{\nu \in \mbb M_1}^{N^2-1} c_\nu e^{i \omega_\nu t} \ha \sigma_\nu}{\sum_{\mu \in \mbb M_1}^{N_\ts{t}-1} c_\mu e^{i \omega_\mu t}}$. 

\emph{Illustration: The stochastic dissipative qubit---} 
Many of the exciting phenomena displayed by non-Hermitian Hamiltonians have been experimentally observed in a dissipative qubit \cite{naghiloo_quantum_2019, chen_quantum_2021, abbasi_topological_NHQubit, erdamar_constraining_2024, Harrington2022, chen_decoherence_2022, wang_dissipative_2023}. The setting corresponds to the Lindblad evolution of a 3-level system $\{\ket{g}, \ket{e}, \ket{f}\}$, where all quantum jumps from $\ket{e}$ to $\ket{g}$ are discarded. This post-selection process leads to the  
non-Hermitian Hamiltonian $\ha H=J \ha \sigma_x - i \Gamma  \hat{\Pi}$, 
where $J$ is the hopping between $\ket{e}$ and $\ket{f}$, 
$\Gamma\geq 0$ is the decay rate from $\ket{e}$ to $\ket{g}$ and  $\ha \Pi = \ket{e}\bra{e}$ denotes the projector over state $\ket{e}$. 
The (shifted) spectrum 
$\varepsilon_\pm + i \frac{\Gamma}{2} = \pm \sqrt{J^2 - \frac{\Gamma^2}{4}}$ is real for  $J\geq\frac{\Gamma}{2}$, and imaginary for $J< \frac{\Gamma}{2}$. These two regimes respectively correspond to (passive) $\mc{PT}$ unbroken and broken symmetry phases \cite{bender_PT_1998}. 

\begin{figure}
    \centering
    \includegraphics[width = \linewidth]{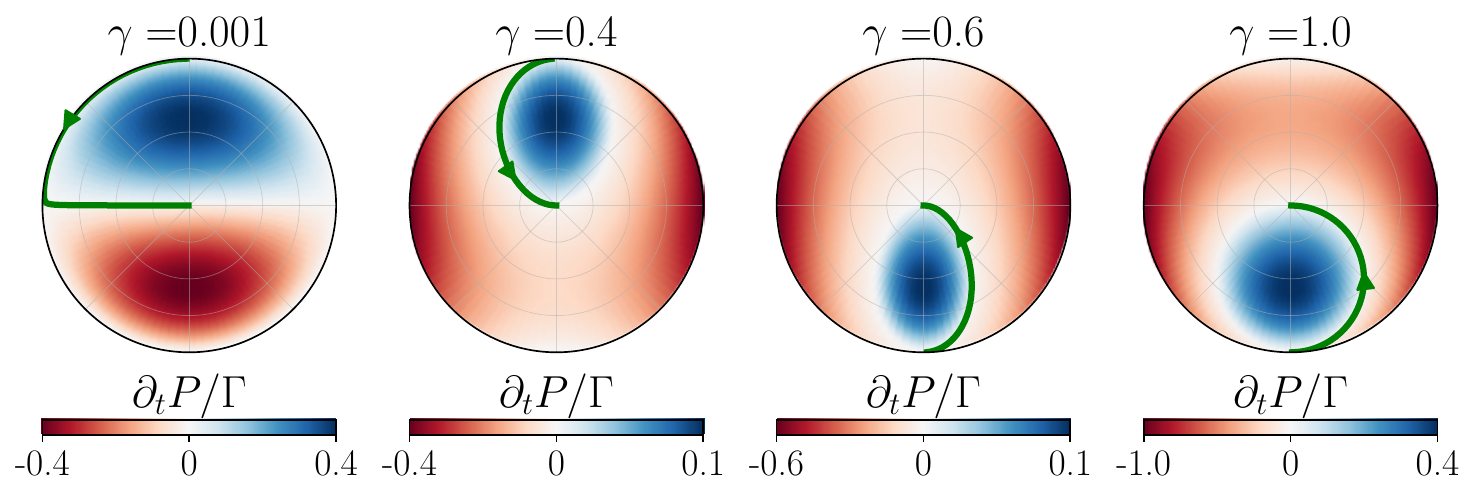}
    \caption{\textbf{Purity evolution} for different strengths of the noise, setting $\Gamma = 1$, as given by \eqref{eq:SDQ_decT} in a cross-section of the Bloch sphere. {The stable steady states corresponding to parameters $(\Gamma/J, \gamma J)$, the hopping being increased in the range $J \in [10^{-2}, 10^2]$ (green line--- See App.~\ref{app:BlochEOMs} for the analytical expressions).}}
    \label{fig:decT_Purity}
\end{figure}

\begin{figure}
    \centering
    \includegraphics[width = \linewidth]{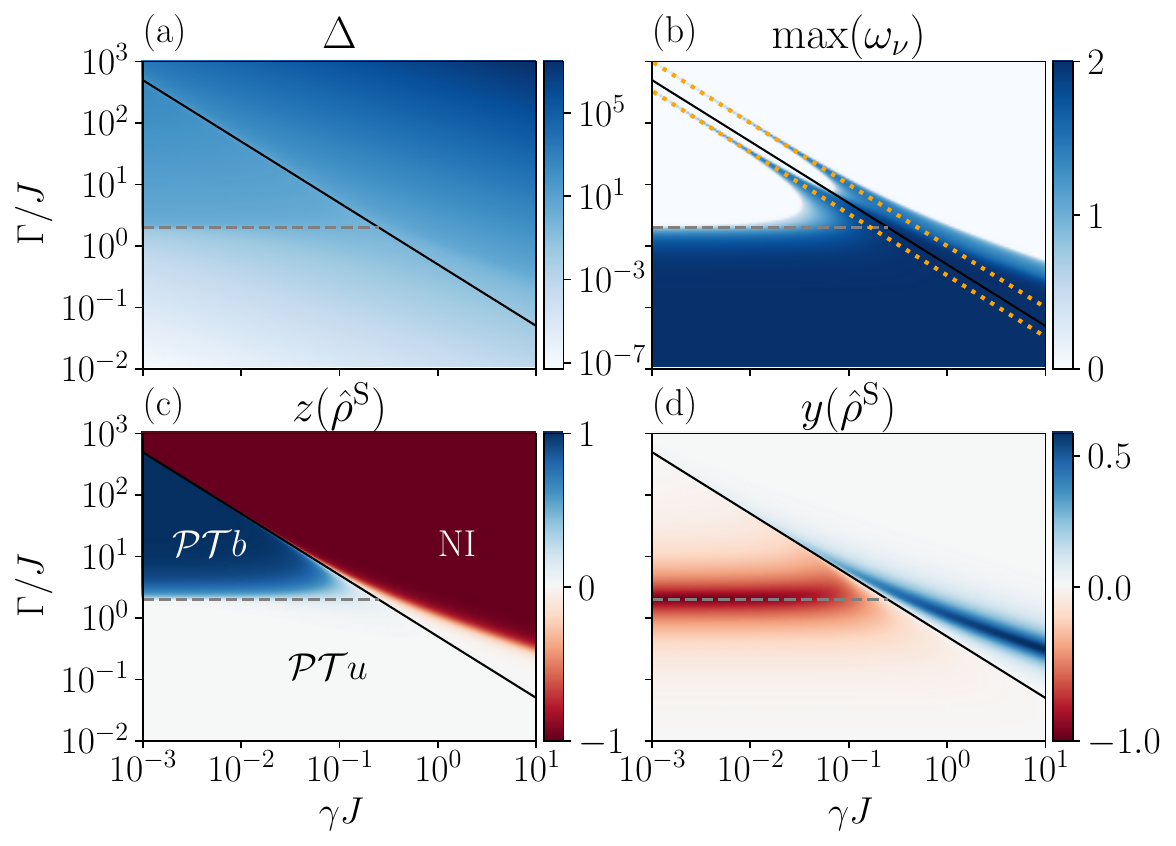}
    \caption{\textbf{Spectral and steady-state phase diagrams}:
    (a) dissipative gap $\Delta$ dictating the time-scale of convergence to the steady-state, (b) maximum imaginary part of the eigenvalues $\omega_\nu$, (c)  $z$ and (d) $y$ components of the Bloch vector for the stable steady state  $\hat{\rho}^{\textsc{s}}$. The $z$ coordinate (c) characterizes the three different phases to which the dynamics can stabilize. The dimensionless parameters determine the noise strength $\gamma J$ and the decay rate $\Gamma/J$. The transition to the noise-induced phase is at $\gamma^* = 1/(2 \Gamma)$ (black line). The transition from $\mc{PT}$ broken to the $\mc{PT}$ unbroken phase happens at $\Gamma/J = 2$ (gray dashed line). In (b) the power laws $\gamma=1/(3 \Gamma)$ and $\gamma = 1/\Gamma$ are also shown (orange dotted lines).}
    \label{fig:PhaseDiags}
\end{figure}

In this work, we are interested in the effect of noise in the anti-Hermitian part of the Hamiltonian. We thus consider a Stochastic Dissipative Qubit (SDQ) with time-dependent Hamiltonian
\begin{align}
    \ha H^\ts{sdq}_t = J\ha \sigma_x - i \Gamma (1+ \sqrt{2 \gamma} \xi_t) \ha \Pi, \label{eq:hamsdq}
\end{align}
which physically corresponds to considering Gaussian fluctuations in the decay parameter $\Gamma$, with strength $\gamma$. {This system can be experimentally realized with superconducting qubits \cite{naghiloo_quantum_2019}, where noise is known to be present in the setup (cf. App. \ref{sec:supercNHQubit}), or in trapped ions \cite{quinn2023observingsuperquantumcorrelationsexceptional}, where noise may be easier to tune (cf. App. \ref{sec:TrapIonNHQUbit}).} The master equation \eqref{BNGL_master_eq} for this system reads
\begin{align} \notag
    \partial_t \ha \rho_t &= - i J [\ha \sigma_x, \ha \rho_t] -(\Gamma - \gamma \Gamma^2)\{ \ha \Pi, \ha \rho_t\} + 2 \gamma \Gamma^2\ha \Pi \ha \rho_t \ha \Pi\\ &\quad + (2 \Gamma - 4 \gamma \Gamma^2) \Tr(\ha \Pi \ha \rho_t) \ha \rho_t. \label{eq:MasterEq_SDQ}
\end{align}

The trace of the 
Liouvillian acting on any state reads $\Tr(\uL[\ha \rho])=(4 \gamma \Gamma^2-2 \Gamma)\Tr(\ha \Pi \ha \rho)$. 
{Note that it depends on the population of the $\ket{e}$ state, since $\Tr(\ha \Pi \ha \rho) = \rho_{ee}$. Therefore the dynamics inhibits the state $\ket{e}$ when $4 \gamma \Gamma^2-2 \Gamma<0$ and favors it when $4 \gamma \Gamma^2-2 \Gamma>0$. Therefore, the dynamics exhibits a transition  at $\gamma^* = \frac{1}{2 \Gamma}$. 
At this critical value, the trace of the Liouvillian vanishes}, the dynamics is Completely Positive, and Trace Preserving (CPTP), and its generator admits the standard GKSL form, with jump operator $\ha \Pi$ and rate $\Gamma$. The dynamics is trivially unitary when the Hamiltonian is Hermitian ($\Gamma=0$). The noise can be used to tune the success rate, i.e., the number of trajectories with no jumps. Assuming that the trace of the unnormalized state gives the success rate, it reads $\Tr(\tilde{\rho}_t)= e^{-2 \Gamma(1-2\gamma \Gamma)t}(c_0 + J B\int_0^t e^{-J B \tau} \rho_{ff}(\tau) \mathrm{d} \tau)$ where $J B= 2 \Gamma(2 \gamma \Gamma-1)$. If the noise increases, the overall decay will be slower; thus, a stronger noise $\gamma$ could be used to mitigate the decay of the success rate, especially near the critical value $\gamma^*$.”

Every state of a qubit can be expressed in the Bloch sphere as $\ha \rho_t = \frac{1}{2}(\ha{\mbb 1} + \be r_t \cdot \ha{\bs \sigma})$, where $\ha{\bs{\sigma}}$ is a vector containing the Pauli matrices 
 \cite{nielsen_quantum_2010}. Expressing the Bloch vector $\be r_t=(x_t, y_t, z_t)$ in spherical coordinates $(r_t, \theta_t, \phi_t)$, the purity \eqref{eq:tauP} evolves for any state $\ha \rho_t$ as 
\begin{equation} \label{eq:SDQ_decT}
  \!\!\!  \partial_t P_t = \Gamma r_t \left[ (2 \gamma \Gamma - 1)(r_t^2-1)\cos\theta_t- \gamma \Gamma r_t \sin^2\theta_t\right]. 
\end{equation}
Note that this expression shows a change in behavior at the TD-TI transition $\Gamma \gamma = \frac{1}{2}$. When $\Gamma\gamma <\frac{1}{2}$, the purifying states are in the north hemisphere of the Bloch sphere, while for $\Gamma\gamma >\frac{1}{2}$, these states are shifted to the southern hemisphere, see Fig. \ref{fig:decT_Purity}. { More details on the steady states are provided in Apps. \ref{app:Liouv_spect}, \ref{app:BlochEOMs}.}

We now investigate the spectral properties of the SDQ Liouvillian \eqref{eq:MasterEq_SDQ}. The phase diagrams in Fig. \ref{fig:PhaseDiags} have three distinct phases: at weak noise $\gamma$, the notions of $\mc{PT}$ broken $\Gamma/J >2 $ and unbroken $\Gamma/J<2 $ symmetry of the NH Hamiltonian govern the dynamics, while at large noise strength, we see a transition to the TI dynamics at $\gamma^*$. We refer to these phases as $\mc{PT}$ broken ($\mc{PT}$b), $\mc{PT}$ unbroken ($\mc{PT}$u) and Noise Induced (NI). Figure \ref{fig:PhaseDiags}(a) shows the dissipative gap $\Delta$ as a function of the noise strength $\gamma$ and the decay rate $\Gamma$. In the $\mc{PT}$u phase, the dissipative gap is very small; thus, the convergence to the stable steady state is very slow. { Interestingly, a small \textit{residual damping rate} was observed experimentally in this phase \cite{naghiloo_quantum_2019} (also see App.~\ref{app:exp_real}).} In the $\mc{PT}$b phase, the gap is larger. Therefore, the steady state is reached faster than in the $\mc{PT}$u phase. In the NI phase, the gap is very large, ensuring fast convergence to the stable steady state [cf. Eq.~\eqref{eq4}]. Note that the gap is always smaller around the transition to the NI phase, which implies that the GKSL dynamics has slower convergence to the stable steady state. Fig. \ref{fig:PhaseDiags}(b) shows the maximum imaginary part of the eigenvalues of the Lindbladian, which dictate the oscillatory behavior of the dynamics. Deep in the $\mc{PT}$b and NI phases, the imaginary part vanishes, implying that the dynamics is not oscillatory. However, it is nonzero in the $\mc{PT}$u phase. This fact, along with the very small dissipative gap, shows what features of the $\mc{PT}$u phase survive the application of classical noise. 

We now characterize the non-degenerate stable steady state $\ha \rho^\ts{s}$. Figure \ref{fig:PhaseDiags}(c) shows the $z$ component of the Bloch vector of $\ha \rho^\ts{s}$, which reads $z(\hat \rho^\ts{s})=-\frac{\lambda_0(\lambda_0-A J)}{4J^2 + \lambda_0 (\lambda_0 - AJ)}$ (cf. App. \ref{app:Liouv_spect}),  where $\lambda_0$ denotes the eigenvalue with the largest real part and $A=\frac{\Gamma}{J}(\gamma \Gamma-1)$ is a constant. In the $\mc{PT}$u phase, the $z$ component is close to zero, and the steady state is close to the maximally mixed state. Thus, when a very small noise is added to the $\mc{PT}$ symmetric NH Hamiltonian,  the time evolution leads to a highly mixed state at very long times. In the $\mc{PT}$b phase, $z$ is close to unity---so the stable steady state is close to $\ket{f}$, as losses induced by $\Gamma$ remove all population in the $\ket{e}$ state. In the NI phase, the steady state is $\ket{e}$---the state that originally leaked to the ground state. This can be interpreted as a noise-induced transition to stability \cite{horsthemke_noise-induced_2006} of the originally unstable state, a feature shared by other noisy dynamics \cite{martinez-azcona_stochastic_2023}. The $y$ Bloch coordinate phase diagram in Fig. \ref{fig:PhaseDiags}(d) shows that the transition from the mixed state to the $\ket{f}$ ($\ket{e}$) state happens by acquiring a negative (positive) value of the $y$ coordinate in the Bloch sphere. The analytic expression for this quantity is $y(\ha \rho^\ts{s})= \frac{2 \lambda_0 J}{4 J^2 +\lambda_0 (\lambda_0 - AJ )}$ (cf. App. \ref{app:Liouv_spect}).

\begin{figure}
    \centering
    \includegraphics[width = .8\linewidth]{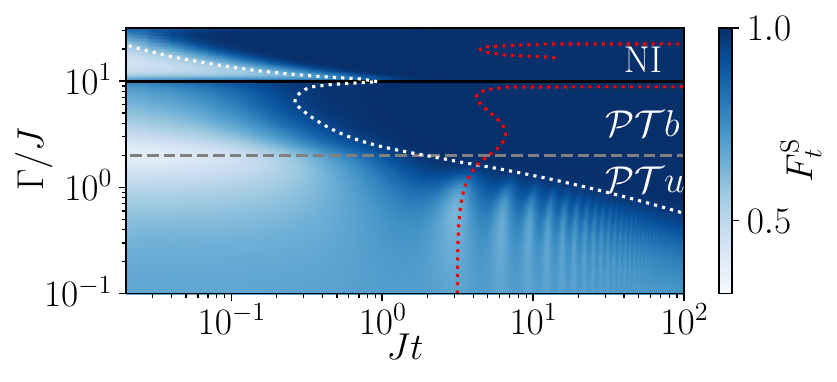}
    \caption{\textbf{Fidelity between the time-evolved state and the stable steady state}
    as a function of the decay parameter $\Gamma$ and time for noise strength $\gamma J = 0.05$. $\mc{PT}$u to $\mc{PT}$b phase transition (gray dashed) and $\mc{PT}$b to NI phase transition (black). Timescales of the dissipative gap $\Delta^{-1}$ (white dotted) and oscillatory dynamics $2 \pi/\omega$ (red dotted). }
    \label{fig:FidSteadyState}
\end{figure}
 
Let us now turn to the dynamics generated by the master equation \eqref{eq:MasterEq_SDQ}.
The dynamics can be solved numerically in two complementary ways (cf. App. \ref{app:BlochEOMs}, \ref{app:Comp_dynamics}). We characterize the distinguishability between the evolving and stable steady states using the Uhlmann fidelity $F^\ts{s}_t=F(\ha \rho_t, \ha \rho^\ts{s})$ \cite{uhlmann_transition_1976, bures_extension_1969}, which admits a compact form in a qubit system \cite{jozsa_fidelity_1994, hubner_explicit_1992} (cf. App. \ref{app:fidelity}).
Figure \ref{fig:FidSteadyState} shows the evolution of the fidelity as a function of time and the decay parameter $\Gamma$ for a small noise, $\gamma J = 0.05$. This value ensures that the three different phases are manifested: (i) for  $\frac{\Gamma}{J}<2$, the system is in the $\mc{PT}$u phase, and the fidelity to the steady state exhibits the anticipated oscillatory behavior until it vanishes at a time close to the inverse dissipative gap $t \sim \Delta^{-1}$ (white dotted line). The period of these oscillations is perfectly characterized by $2 \pi/\omega$  where $\omega = \max_\nu \mr{Im}(\lambda_\nu)$, as highlighted by the red dotted line. Note that this oscillatory frequency is affected by the increase of $\Gamma$ as we approach the $\mc{PT}$ phase transition, as also observed experimentally in the dissipative qubit \cite{naghiloo_quantum_2019}. (ii) For $2 <\frac{\Gamma}{J} < \frac{1}{2 J \gamma}$, the system is in the $\mc{PT}$b phase, with a larger dissipative gap [cf. Fig. \ref{fig:PhaseDiags}(a)], and thus a faster convergence to the steady state. Interestingly, the oscillatory timescale can be finite outside the $\mc{PT}$u phase as $\omega$ does not vanish identically. In particular, it shows minima at $\Gamma = \frac{1}{3 \gamma}$ and $\Gamma = \frac{1}{\gamma}$, as observed in Fig. \ref{fig:PhaseDiags}(b). Exactly at the transition $\Gamma = \frac{1}{2 \gamma}$, the dynamics is CPTP, and its convergence is slower, as explained by the smaller gap at the transition line. (iii) For $\Gamma>\frac{1}{2 \gamma}$, the system is in the NI phase and exhibits the fastest convergence to the steady state. Experimentally, the decay rate is $\Gamma = 6.7 \; \mu \mr{s}^{-1}$ \cite{naghiloo_quantum_2019}, which upper bounds the noise strength $\gamma < \gamma^* \approx 0.074 \;  \mu \mr{s}$ given that no NI phase was observed.

\emph{Conclusions---} 
We have considered the time evolution governed by a fluctuating non-Hermitian Hamiltonian describing a quantum mechanical system subject to stochastic 
 gain and loss. The resulting noise-averaged dynamics is described by a novel anti-dephasing master equation beyond the GKSL form.
We characterized the purity dynamics and found that the stable steady states live in the Liouvillian eigenspace whose eigenvalues have the largest real part. The salient features of such dynamics are manifested in a stochastic dissipative qubit. The addition of noise allows for control over the steady states and the convergence rate. We find three main dynamical phases: the $\mc{PT}$ unbroken and broken regimes, complemented with a noise-induced phase.  
Our results are amenable to current experimental platforms realizing non-Hermitian evolution and provide a framework to capture the effect of noise, as exemplified by the residual damping rate \cite{naghiloo_quantum_2019}. Our findings suggest that engineering fluctuating operators associated with gain and loss may open new avenues for quantum state preparation \cite{verstraete_quantum_2009}. Our results may also be applied to understand the stability of imaginary time evolution \cite{motta_determining_2020} against classical noise.

\textit{Acknowledgements}---
We thank Oskar Pro\'sniak, Kohei Kawabata, Martin Plenio, Raphaël Chétrite, and Yogesh Joglekar for their insightful discussions and Juan Garrahan and Shiue-Yuan Shiau for comments on the manuscript. 
This research was funded by the Luxembourg National Research Fund (FNR, Attract grant QOMPET 15382998 and grant 17132054). The work at Los Alamos was supported by the U.S. Department of Energy.


%

\onecolumngrid

\appendix

\section{Gauge transformations of the anti-dephasing master equation}\label{app:gauge}
Below, we discuss how trace-increasing dynamics can be changed to conserve or reduce the state trace. 
For this, we explicitly distinguish the stochastic $\ha L_\ts{s}$ and deterministic $\ha L_\ts{d}$ parts of the anti-Hermitian part of the Hamiltonian, and add an imaginary deterministic shift $a \in \mbb R$. Namely, we consider 
\begin{equation}
	\ha H_t' = \ha H_0 - i (\ha L_\ts{d} + a\, \ha{\mathbb{1}}) - i \sqrt{2 \gamma }\xi_t \ha L_\ts{s} ,
\end{equation}
which generates the nonlinear master equation 
\begin{align} 
	\partial_t \ha \rho_t = &- i [\ha H_0, \ha \rho_t] - \{\ha L_\ts{d} + a \, \ha{\mathbb{1}}, \ha \rho_t\} + \gamma \{\ha L_\ts{s} , \{ \ha L_\ts{s} , \ha \rho_t\}\} + 2 \Tr\big((\ha L_\ts{d}+ a \,\ha{\mathbb{1}}) \ha \rho_t\big) \ha \rho_t - 4 \gamma \Tr(\ha L_\ts{s}^2 \ha \rho_t)\ha \rho_t.
\end{align}
Systems with any shift $a$ have the same dynamics, since the terms $- \{a \, \ha{\mathbb{1}}, \ha \rho_t\}$ and $+ 2 \Tr(a \ha \rho_t) \ha \rho_t$ exactly cancel each other. This provides a non-Hermitian generalization of the well-known fact that shifting the Hamiltonian by a constant $\ha H + a \mbb{1}$ has no consequence on the dynamics. Thus, a complex zero of energy also has no dynamical effect, provided that the trace of the density matrix is renormalized. It follows that the dynamics can be made TD or even TP with a suitable choice of $a$, namely 
\begin{equation}
	4 \gamma \Tr(\ha L^2_\ts{s} \ha \rho_t) -2 \Tr (\ha L_\ts{d} \ha \rho_t)  - a \leq 0,
\end{equation}
where `$<$' corresponds to a TD map and `$=$' to TP dynamics. 

For the SDQ considered in the main text, $\ha L_\ts{d}=\ha L_\ts{s}=\Gamma \hat{\Pi}$, so the shift has to obey the condition
\begin{equation}
	a \geq (4 \gamma \Gamma^2 - 2 \Gamma) \Tr(\ha \Pi \ha \rho),  
\end{equation}
for any $\ha \rho$. Since the expectation value has the property $\Tr(\ha \Pi \ha \rho )= \rho_{ee}\leq 1$, a looser but state-independent condition on the shift reads  
\begin{equation}
	a \geq 4 \gamma \Gamma^2 - 2 \Gamma= B J.
\end{equation}

Alternatively, consider a gauge transformation that shifts the stochastic component by a constant $b \in \mbb R$
\begin{equation}
	\ha H_t' = \ha H_0 - i \ha L'_\ts{d} - i \sqrt{2 \gamma }\xi_t (\ha L_\ts{s}+ b\ha{\mbb{1}}) ,
\end{equation}
which generates the nonlinear master equation 
\begin{align} 
	\partial_t \ha \rho_t = &- i [\ha H_0, \ha \rho_t] - \{\ha L'_\ts{d} - 4 \gamma b \ha L_\ts{s}, \ha \rho_t\} + \gamma \{\ha L_\ts{s}, \{ \ha L _\ts{s}, \ha \rho_t\}\} + 2 \Tr((\ha L'_\ts{d}-4 \gamma b \ha L_\ts{s}) \ha \rho_t) \ha \rho_t - 4 \gamma \Tr(\ha L^2_\ts{s} \ha \rho_t)\ha \rho_t.
\end{align}
It follows that the transformation 
\begin{equation}
	\begin{cases}
		\ha L_\ts{s} \rightarrow \ha L_\ts{s}+b \, \ha{\mbb{1}},\\
		\ha L_\ts{d} \rightarrow \ha L_\ts{d} + 4 \gamma b \, \ha L_\ts{s},
	\end{cases}
\end{equation}
also leaves the nonlinear master equation invariant. This transformation, similar to the GKSL case \cite{BreuerBook}, allows choosing jump operators $\ha L_\ts{s}$ that are traceless.

\section{Experimental Realization of Stochastic Non-Hermitian Hamiltonians}
\label{app:exp_real}
{
	In this Section, we discuss possible experimental platforms for realizing stochastic non-Hermitian Hamiltonians and analyze their advantages and disadvantages.
	We begin by reviewing the experimental realization of the non-Hermitian qubit in superconducting circuits \cite{naghiloo_quantum_2019}. In particular, we focus on how noise affects this platform. Then, to show that noise in the anti-Hermitian part of the Hamiltonian does not need to be postulated \textit{ad hoc} but can be obtained from first principles, we review the formalism by Carmichael \cite{carmichael_open_1993} in homodyne detection of the degenerate parametric oscillator. Lastly, since tuning the strength of the noise in both of these examples seems quite challenging, we review the setup of the trapped ion non-Hermitian qubit \cite{quinn2023observingsuperquantumcorrelationsexceptional}, which may allow for a tunable noise.
}

\subsection{Noise in the superconducting non-Hermitian qubit}
\label{sec:supercNHQubit}
{\aritra
	The superconducting non-Hermitian qubit studied in \cite{naghiloo_quantum_2019} is built from the three lowest levels of a transmon circuit, which is embedded in a cavity with an impedance mismatch element that allows tuning its density of states, thus modifying the decay of the transmon. In addition, an external magnetic flux is threaded through the SQUID loop of the transmon to tune the decay parameter $\Gamma$ of the effective non-Hermitian system. A Josephson parametric amplifier is used for state tomography and post-selection of quantum jumps.  
	
	The ideal (noise-less) behavior of the non-Hermitian qubit is such that, in the $\mc{PT}$ unbroken phase, its Hamiltonian's eigenenergies are purely real and thus describe a purely oscillatory behavior. In contrast, in the $\mc{PT}$-broken phase, the eigenvalues become purely imaginary, describing the system states as decaying with a real exponential. However, the experiments show \cite{naghiloo_quantum_2019}, in the $\mc{PT}$ unbroken phase, a residual small damping term $\Gamma_\mr{R}$ describing a damped oscillation, which is larger than expected from the decay channel of the $\ket{f}$ state. The authors interpreted this residual decay rate to be associated with the ``\textit{charge and flux noise}'' \cite{naghiloo_quantum_2019}. Our model brings some theoretical understanding of the flux noise. 
	
	As explained before, the magnetic flux tunes the non-Hermitian decay rate $\Gamma$, such that noise in the flux corresponds to noise in the anti-Hermitian part. Therefore, our SDQ models flux noise acting on the NH qubit. This noise is unavoidable in the experimental setup. Unfortunately, this also implies that tuning noise in this setup would be difficult. Still, our results show that the $\mc{PT}$ unbroken phase in Fig. 2(a) has a small but nonzero, dissipative gap. So, our formalism can serve as a minimal model to capture the effect of noise on the experimental setup. 
}

\subsection{The degenerate parametric oscillator}\label{app:degParOsc}
{
	Carmichael derived a stochastic non-Hermitian Hamiltonian for a model of homodyne detection of the degenerate parametric oscillator \cite{carmichael_open_1993}. We briefly review some of its most important features, along with some extensions, for completeness. With this, we aim to show that the stochasticity, particularly noise in the anti-Hermitian part of the Hamiltonian, does not need to arise from uncontrolled interactions with an environment, but can come from a formal treatment of the measurement process. 
	
	The physical setup consists of two optical cavities $A$ and $B$ with respective annihilation operators $\ha a, \; \ha b$. The cavity $\ha a$ with frequency $\omega$ is prepared in the vacuum state $\ket{0}$. Carmichael's setup \cite{carmichael_open_1993} contains a nonlinear crystal, parametrized by $\lambda$, and radiates squeezed light. However, this parameter is unnecessary to understand the main idea so that we will set it to zero $\lambda = 0$. Cavity $\ha b$ is prepared in a coherent state $\ket{\beta}$ and radiates the local oscillator field. Both cavities are leaky and thus modeled by a quantum optical Lindblad master equation with rates $2 \kappa$ for the pumped cavity and $2 \gamma$ for the local oscillator. The two output fields go through a beam splitter, with a reflection coefficient $R$, before arriving at the detector. The evolution is then split into two different steps: a non-unitary evolution corresponding to no quantum jumps at the detector and a collapse when a photoelectron is emitted.
	
	The model can be simplified to get a pure state evolution of the cavity A, conditioned on the measurement output, given by $\ket{\psi^{(a)}_c(t)}$. To do so, the reflectivity and the local oscillator decay rate are sent to zero ($R \rightarrow 0, \; \gamma \rightarrow 0$) while the local oscillator amplitude is taken to infinity $\beta \rightarrow \infty$, to keep the local oscillator photon flux constant, $f = 2 \gamma R |\beta|^2$. Under these conditions, the unnormalized conditioned state $\ket{\ti \psi_c^{(a)}}$ evolves with the non-Hermitian Hamiltonian 
	\begin{equation}
		\ha H = \hbar(\omega - i \kappa) \ha a\dg \ha a - i\hbar \sqrt{2 \kappa f} e^{- i \theta} e^{i \omega t} \ha a,
	\end{equation}
	and is interrupted by collapses 
	\begin{equation}
		\ket{\ti \psi_c^{(a)}} \rightarrow \ha C \ket{\ti \psi_c^{(a)}} = \left(\sqrt{f}e^{i (\theta- \omega t)} + \sqrt{2 \kappa} \ha a \right )\ket{\ti \psi_c^{(a)}},
	\end{equation}
	where $\theta$ is the local oscillator phase. The collapses happen at probability $p_c(t)~=~\delta t{\braket{\ti \psi_c^{(a)}(t)|\ha C\dg \ha C|\ti \psi_c^{(a)}(t)}}/{\braket{\ti \psi_c^{(a)}(t)|\ti \psi_c^{(a)}(t)}} $. Taking the limit $f/(2 \kappa) \rightarrow \infty$, the wavefunction undergoes an infinite number of infinitesimal collapses in any time interval, such that the jump process reduces to white noise. After some algebra, taking the time between successive collapses as $\tau_{n+1} \sim f^{-1}$ and the total number of collapses $m$ in the increment of time $\delta t$ from a Gaussian distribution \cite{carmichael_open_1993}, we find that the conditioned, unnormalized state of the source $\ket{\ti \psi_c^{(a)}(t)}$ evolves with a Schrödinger equation
	\begin{equation}
		i \hbar \partial_t \ket{\ti \psi_c^{(a)}(t)} = \ha H_\xi(t) \ket{\ti \psi_c^{(a)}(t)},
	\end{equation}
	where $\ha H_\xi(t)$ is a stochastic, non-Hermitian Hamiltonian given by  
	\begin{align}
		\ha H_\xi(t) &= \hbar(\omega - i \kappa) \ha a\dg \ha a + i \hbar\left( \sqrt{2 \kappa} \braket{\psi_c^{(a)}(t)|\ha X_\theta|\psi_c^{(a)}(t)} + \xi_t\right) e^{- i \theta} \sqrt{2 \kappa} \ha a, 
	\end{align}
	where we introduced the operator $\ha X_\theta = e^{i \theta} \ha a\dg + e^{- i \theta} \ha a$ and $\xi_t$ is a Gaussian white noise. 
	Note that Carmichael \cite{carmichael_open_1993} uses the interaction picture with respect to $\omega \ha a\dg \ha a$, leading to a purely anti-Hermitian  Hamiltonian. Here, for clarity, we do not use the interaction picture and thus obtain a Hamiltonian with both a Hermitian and an anti-Hermitian part. Interestingly, a similar stochastic non-Hermitian Hamiltonian has been proposed by Pi\~nol \textit{et al.} \cite{pinol_diff_unravelings_2024} for a two-level system instead of an oscillator, i.e. with $\ha \sigma_\pm$ instead of $\ha a, \ha a\dg$.
}

\subsection{Trapped Ion realization}
\label{sec:TrapIonNHQUbit}
{
	Recently, a non-Hermitian trapped-ion qubit has been realized \cite{quinn2023observingsuperquantumcorrelationsexceptional}. The experimental realization uses a single $ ^{40}\mathrm{Ca}^+$ ion in a linear-Paul trap \cite{sherman_experimental_40Ca_2013}. The two energy levels used to build a qubit are $\ket{\uparrow}= \ket{m = +5/2}$ and $\ket{\downarrow}= \ket{m=+3/2}$ within the meta-stable $D_{5/2}$ manifold. The hopping between the states $J\sigma_x$ uses resonant radio frequency pulses at the qubit frequency. To engineer the decay, the state $\ket{\downarrow}$ is coupled to the short-lived $P_{3/2}$ state $\ket{A}$, which decays primarily to the $S_{1/2}$ ground state with rate $\gamma_g$. The coupling to the $\ket{A}$ state is achieved with $\pi$-polarized light of pulse strength $J_A$. For $\gamma_g\gg J_A$, the auxiliary level $\ket{A}$ may be adiabatically eliminated \cite{Brion_2007, Muga_2008} and through post-selection, an effective Non-Hermitian Qubit is obtained for the levels $\{\ket{\downarrow}, \ket{\uparrow}\}$. The effective decay rate of $\ket{\downarrow}$ is then $\gamma = J_A^2/\gamma_g \ll \gamma_g$. The pulse strength may then be modulated in time with some \textit{external} and \textit{tunable} white noise $J_A (1 + \sqrt{2 \gamma'}\xi_t')$, to achieve a fluctuating decay rate. 
	
	A limitation of the experimental setup is that $5.87\%$ of the population of the $P_{3/2}$ state decays back to the $D_{5/2}$ manifold \cite{quinn2023observingsuperquantumcorrelationsexceptional}, limiting the time that the effective non-Hermitian description is valid over one or two Rabi oscillations. This might still be a suitable platform to observe our predicted results. Indeed, one of the main results of the Stochastic Dissipative Qubit is the presence of a noise-induced phase. Figure 3 of the main text shows that the dissipative gap time scale in the NI phase is much shorter than the Rabi oscillation time, which implies that the NI phase should be experimentally observable in this trapped-ion platform.
}

\section{Liouvillian spectrum of the Stochastic Dissipative Qubit}\label{app:Liouv_spect}
The master equations given in the main text can be formulated in Liouville superspace. To this end, the density operator is written as a vector $|\tilde{\rho})$ of the form
\begin{equation}
	\urho = \sum_{n,m} \urho_{nm} \ket{n}\bra{m} \; \mapsto \;  |\urho) = \sum_{n,m} \urho_{nm} \ket{n} \otimes \ket{m}^*;
\end{equation}
we refer the reader to \cite{Gyamfi_2020} for a detailed description of this procedure. 
Superoperators are mapped to operators on the superspace following the \textit{Choi-Jamiołkowski} isomorphism \cite{CHOI1975285, JAMIOLKOWSKI1972275}
\begin{equation}
	{\hat{X}} \urho {\hat{Y}} \; \mapsto \;  ({\hat{X}} \otimes {\hat{Y}}^\intercal)|\urho),
\end{equation}
where $\otimes$ denotes the Kronecker product and $\bullet^\intercal$ the transpose. 

Vectorization requires fixing the inner product to be taken between operators; in the above procedure, the inner product is chosen as the standard \textit{Hilbert-Schmidt} inner product $(X|Y)= \Tr(\hat{X}^\dagger \hat{Y})$, which for vectorized operators conveniently reduces to the standard Euclidean inner product for vectors, $(X|Y) = \sum_{m,n} X_{mn}^* Y_{mn}$.

The non-trace-preserving state dynamics is then dictated by $\frac{\mr d}{\mr dt}|\tilde{\rho}) = \tilde{\mathcal{L}}|\tilde{\rho})$ with the vectorized Liouvillian superoperator in equation (3) of the main text given by
\begin{align}
		\tilde{\mathcal{L}}= -i (\ha H_0 \otimes \ha{\mathbb{1}} - \ha{\mathbb{1}} \otimes \ha H_0^\intercal) - (\ha L \otimes \ha{\mathbb{1}} + \ha{\mathbb{1}} \otimes \ha L^\intercal)+ \gamma (\ha L^2 \otimes \ha{\mathbb{1}} + \ha{\mathbb{1}} \otimes (\ha L^2)^\intercal + 2 \ha L \otimes \ha L^\intercal).
\end{align}
Similarly, the Liouvillian for the SDQ (cf. eq. (8) main text) is given in matrix form by
\begin{equation} \label{eqSM:L}
	\uL = J \left( \begin{array}{cccc}
		0 & i & - i & 0\\
		i & A & 0 & -i \\
		- i & 0 & A & i \\
		0 & - i & i & B
	\end{array}\right),
\end{equation}
where we have defined the constants $A = \frac{\Gamma}{J}( \gamma \Gamma-1)$ and $B=2\frac{\Gamma}{J}(2 \gamma \Gamma-1)$.
The characteristic polynomial of the Liouvillian thus reads 
\begin{equation}
	(A- \Lambda)f(\Lambda)=0,
\end{equation}
with the  cubic polynomial $f(\Lambda) = \Lambda^3 - \Lambda^2 (A+B)+ \Lambda (4 + AB)- 2B$. 
To find its roots, we first shift the variable, using $z = \Lambda - (A+B)/3$ and get a \textit{depressed} cubic, i.e. a cubic equation without the quadratic term 
\begin{equation}
	f(z) = z^3 + 3 C z + D = 0,
\end{equation}
with the constants $C= - \big(\frac{A+B}{3}\big)^2 + \frac{4+AB}{3}$ and $D = \frac{A+B}{3}(4+AB) - 2B - 2 \big(\frac{A+B}{3}\big)^3$. We then use Cardan's trick \cite{Nickalls_1993}, also known as Vieta's substitution, that is, set $z= U + \frac{a}{U}$, choosing $a$ to remove the terms in $U$ and $1/U$, i.e., $a=-C$. This leads to a quadratic equation in $U^3$, 
\begin{equation}\label{eq:quadratic_u3}
	(U^3)^2 + D U^3 - C^3=0,
\end{equation}
with solutions $U_{m,\pm}=e^{m i \frac{2\pi}{3}}\big[-\frac{D}{2}\pm \sqrt{\big(\frac{D}{2}\big)^2+ C^3}\big]^{1/3}$, with $m=(0,\pm 1)$. 
The Liouvillian eigenvalues follow as 
\begin{equation}
	\Lambda_{m,\pm}=J \left( U_{m,\pm}-\frac{C}{U_{m,\pm}}+ \frac{A+B}{3}\right).
\end{equation}

Note that we seem to have obtained six solutions from a cubic equation. However, three pairs of solutions are the same. To show this, let $V= - \frac{D}{2}\pm \sqrt{\frac{D^2}{4}+C^3} \equiv U_{m,\pm}^3$ denote any nonzero root of the quadratic equation \eqref{eq:quadratic_u3}. If $V$ is a root then $\frac{-C^3}{V}$ is also a root; this implies that if we change $+ \rightarrow -$ in $\Lambda_{m, \pm}$ we exchange the terms $U_{m, \pm}$ with the term $-\frac{C}{U_{m,\pm}}$. So it is enough to consider the eigenvalues $\Lambda_{m,+}$.

We denote the eigenvalues $\{\lambda_0\equiv \Lambda_{0,+}, \lambda_1\equiv \Lambda_{1,+}, \lambda_2\equiv \Lambda_{-1,+}, \lambda_3=AJ\}$ that diagonalize the Liouvillian \eqref{eqSM:L}, 
\begin{equation}
	\uL = \sum_{\nu=0}^3 \lambda_\nu |B_\nu)(B_\nu|.
\end{equation}

\begin{figure}[h]
	\centering
	\includegraphics[width = .4\linewidth]{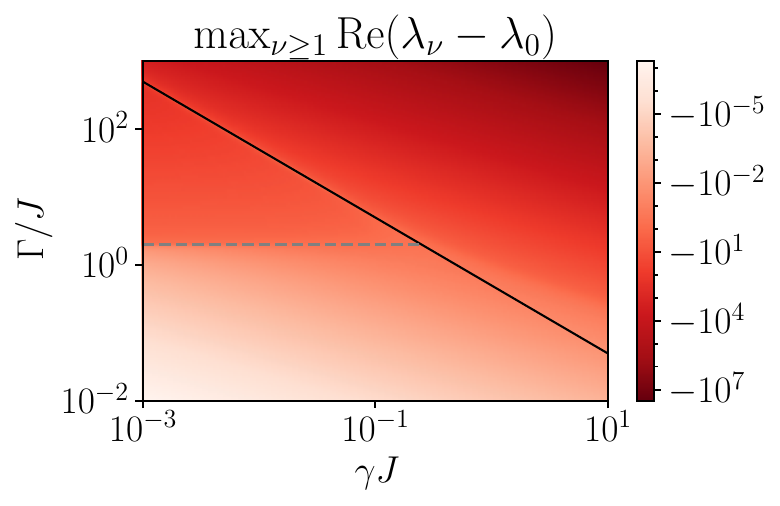}
	\caption{Maximum difference of the real part of $\lambda_\nu, \nu\geq 1$ with $\lambda_0$. The difference is always negative for the displayed range of parameters. Therefore, $\lambda_0$ is always the eigenvalue with the largest real part. The transitions between the $\mc{PT}$u, $\mc{PT}$b, and NI phases are shown in the dashed and solid line as in Fig. 2 of the main text. }
	\label{fig:diff_evals}
\end{figure}

\begin{figure}[h]
	\centering
	\includegraphics[width = .65 \linewidth]{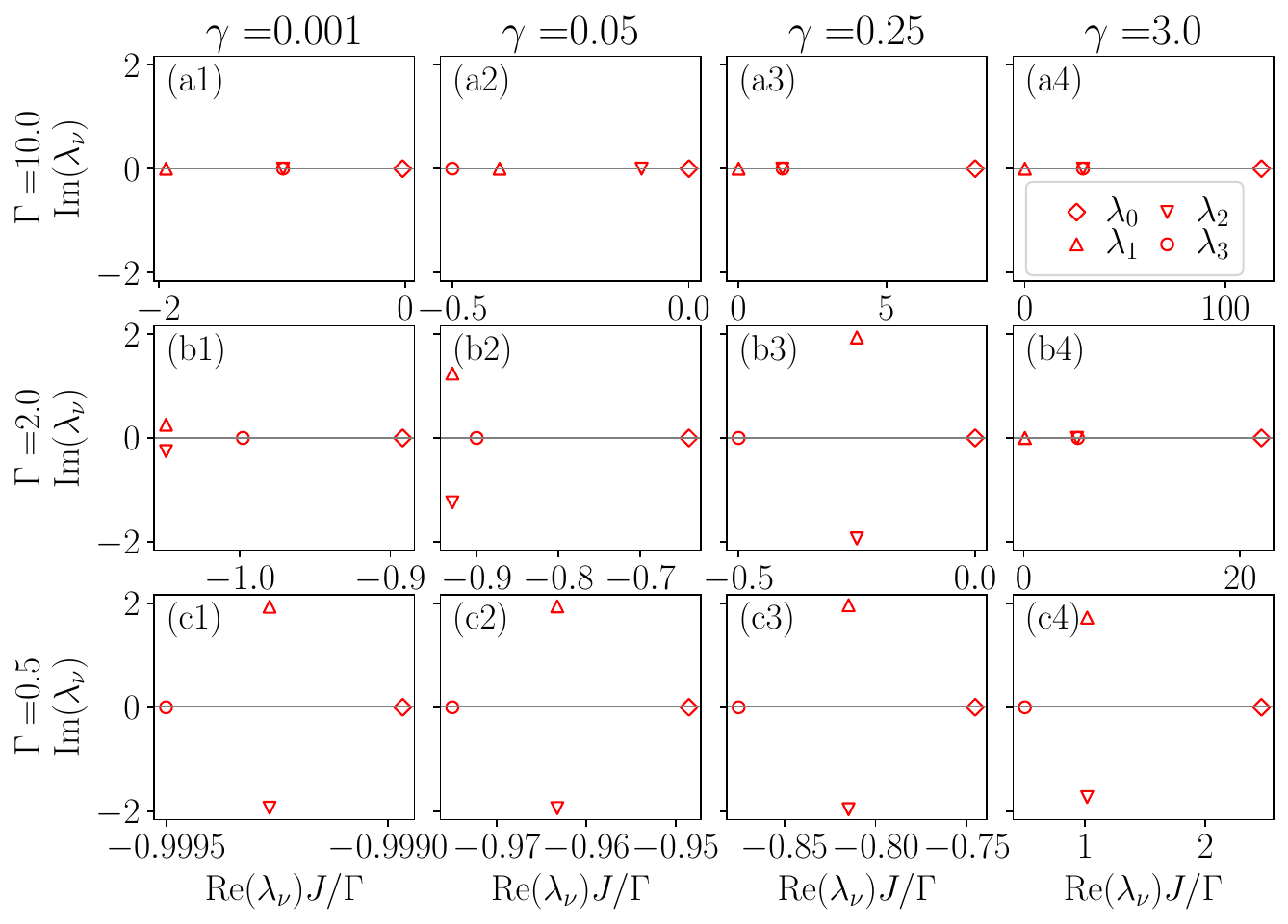}
	\caption{Spectrum of the SDQ Liouvillian for different locations in the phase diagram of the system. The parameters span the different phases: (a1) $\mc{PT}$ broken,  (c1-3) $\mc{PT}$ unbroken, and (a3, a4, b4) Noise Induced;  as well as the transitions between them: (b1,2) $\mc{PT}$ breaking transition, (a2,b3) TD-TI transition and (c4) transition from mixed state to $\ket{e}$ state. Note that the $x$-axis has been rescaled by $\Gamma/J$.}
	\label{fig:SDQ_spectrum}
\end{figure}

In the domain of interest, with  $\gamma$ and $\Gamma$ positive, $\lambda_0$ is the eigenvalue with the maximum real part. This feature is verified for a wide range of parameters in Fig. \ref{fig:diff_evals}, where the maximum difference of the real part of eigenvalues $\lambda_\nu$ for $\nu\geq 1$ and $\lambda_0$ is always shown to be negative. The full spectrum can be checked for many different parameters in Fig. \ref{fig:SDQ_spectrum}. This plot allows us to see the large dissipative gap and no oscillatory behavior in the NI phase (a3, a4, b4), as well as the nonzero imaginary components and a small gap in the $\mc{PT}$u phase (c1-3), and the intermediate gap and no oscillations in the $\mc{PT}$b phase (a1), in addition to the transitions between them.

The eigenvector $|B_0)= (b_0^{(0)}, b_1^{(0)}, b_2^{(0)}, b_3^{(0)})^\intercal$ associated to the eigenvalue $\lambda_0$  is given by the solution of the system of equations
\begin{equation}
	\begin{cases}
		+i (b^{(0)}_1-b^{(0)}_2) = \lambda_0 b^{(0)}_0,\\
		+i (b^{(0)}_0-b^{(0)}_3)=(\lambda_0-A)b^{(0)}_1,\\
		-i (b^{(0)}_0-b^{(0)}_3)=(\lambda_0-A)b^{(0)}_2,\\
		-i (b^{(0)}_1-b^{(0)}_2)=(\lambda_0-B)b^{(0)}_3.\\
	\end{cases} 
\end{equation}
By substitutions, given that $\lambda_0$ is the largest eigenvalue, in particular $\lambda_0 > A$, 
we find 
\begin{equation}
	b^{(0)}_3 = b^{(0)}_0 \left( 1 + \lambda_0 \frac{\lambda_0 - A}{2} \right),
\end{equation}
and $b^{(0)}_1=-b^{(0)}_2= - i b^{(0)}_0 \lambda_0/2$. We choose $b^{(0)}_0$ such that the eigenvector is normalized for real $\lambda_0$, yielding the eigenvector associated to the stable steady state 
\begin{equation}\label{eq:steadystate_analyt}
	|B_0) = \frac{1}{2 + \lambda_0\frac{\lambda_0-A}{2}} \left(\begin{array}{c}
		1\\
		-\frac{i}{2}\lambda_0\\
		\frac{i}{2}\lambda_0\\
		1 + \lambda_0\frac{\lambda_0 - A}{2}
	\end{array}\right).
\end{equation}
\section{Bloch sphere dynamics for the Stochastic Dissipative Qubit}\label{app:BlochEOMs}
Any density matrix of a qubit $\hat{\rho}$ is completely characterized by its Bloch coordinates $\be r = (x,y,z)$ from the decomposition 
\begin{equation}
	\ha \rho = \frac{1}{2}(\ha{\mathbb{1}} + \be r \cdot \ha{\bs \sigma}),
\end{equation}
where $\ha{\bs \sigma}$ is a vector containing the Pauli matrices obeying the standard commutation $[\ha \sigma_{n}, \ha \sigma_{m}]= 2 i \epsilon_{nml} \ha \sigma_l$, and anticommutation $\{\ha \sigma_n, \ha \sigma_m\}= 2 \delta_{nm} \ha{\mbb 1}$ relations. 

For the SDQ, the master equation (4) in the main text dictates the evolution of these coordinates according to the coupled differential equations
\begin{equation}\label{eq:EOM_bloch}
	\begin{cases}
		\dot x = -(\gamma \Gamma^2 + z \Gamma( 1 - 2 \gamma \Gamma))x, \\
		\dot y = - 2 J z -(\gamma \Gamma^2 + z \Gamma( 1 - 2 \gamma \Gamma))y, \\
		\dot z =  2 J y -\Gamma( 1 - 2 \gamma \Gamma)(z^2-1).
	\end{cases}
\end{equation}

\begin{figure}
	\centering
	\includegraphics[width = .75 \linewidth]{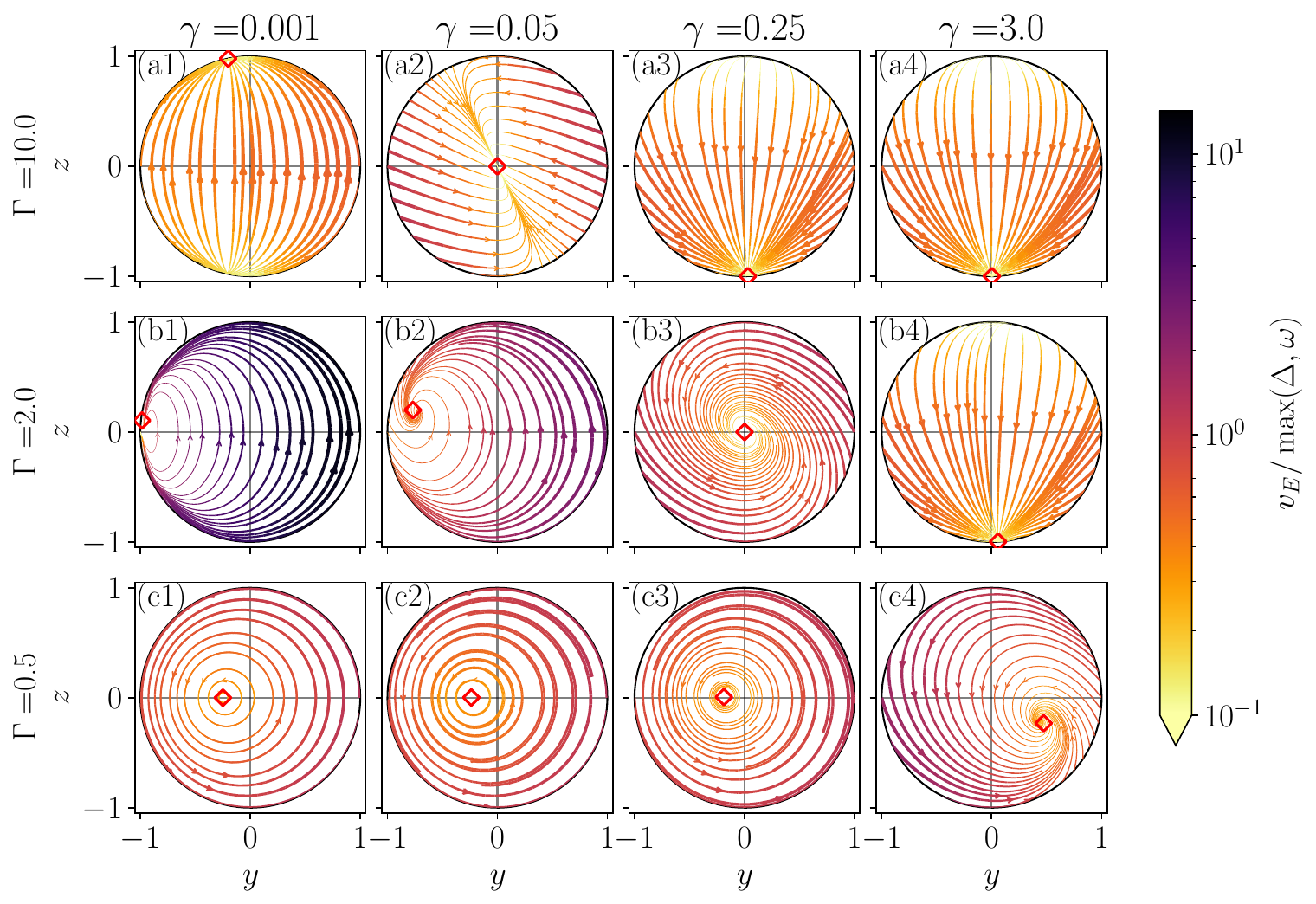}
	\caption{Streamlines of the vector fields in the cross-section of the Bloch sphere with $x=0$. The parameters span the different phases: (a1) $\mc{PT}$ broken,  (c1-3) $\mc{PT}$ unbroken, and (a3, a4, b4) Noise Induced;  as well as the transitions between them: (b1,2) $\mc{PT}$ breaking transition, (a2,  b3) TD-TI transition and (c4) transition from mixed state to $\ket{e}$ state. The analytical steady state \eqref{eq:steadystate_analyt} is shown as the red diamond. The color and linewidth represent the Euclidean speed $v_E = \sqrt{(\dot y)^2 + (\dot z)^2}$, divided by the maximum of the two main frequency units: the dissipative gap $\Delta$ and the maximum imaginary part of the eigenvalues $\omega= \max_\nu(\mr{Im}(\lambda_\nu))$.}
	\label{fig:streamlines}
\end{figure}

The streamlines of this vector field, as well as the analytical steady state $|B_0)$ \eqref{eq:steadystate_analyt} are compared in Fig. \ref{fig:streamlines}. We see perfect agreement between the analytical steady state (red diamond) and the evolution of the streamlines showing convergence to it. 

The behavior discussed in the main text is also apparent from these coordinates: (i) the $\mc{PT}$b phase has a steady state in the north pole, very close to the $\ket{f}$ state [cf. Fig. \ref{fig:streamlines}(a1)]; (ii) the $\mc{PT}$u phase [cf. Fig. \ref{fig:streamlines}(c1-3)] has a steady state close to the maximally mixed state in the center of the sphere, with a small $y$ component, as observed in Fig. 2(d) of the main text and (iii) the NI phase [cf. Fig. \ref{fig:streamlines}(a3-4, b4)] has a steady state very close to the $\ket{e}$ state in the south pole of the sphere. 

The transition between the different phases also exhibits an interesting behavior: In the $\mc{PT}$u to $\mc{PT}$b phase transition (cf. Fig. \ref{fig:streamlines} (b1-2)), we see a convergence to a state close to $\ket{-y}= \frac{\ket{f}-i\ket{e}}{\sqrt{2}}$, interestingly, the color scale shows that the speed of this convergence is larger than the dissipative gap or the oscillatory timescale; the transition from TD to TI dynamics is also interesting (cf. Fig. \ref{fig:streamlines} (a2, b3)), in this transition the dynamics is CPTP, and the steady state is the maximally mixed state in the center of the Bloch ball, to which the convergence shows almost no oscillations in (a2) and an oscillatory behavior in (b3), see Fig. \ref{fig:SDQ_spectrum} to understand this. Lastly, the transition from $\mc{PT}$u to the NI phases (cf. Fig. \ref{fig:streamlines} (c4)) shows an oscillatory convergence to the steady state, which has a positive component of $y$, as already known from Fig. 2 (d).

\subsection{Polar coordinates: Nullclines}
{
	The Bloch sphere can be naturally parametrized in spherical coordinates $(r, \theta, \phi)$, related to the $(x,y,z)$ variables by the standard relations $x = r \sin \theta \cos \phi, \; y = r \sin \theta \sin \phi, \; z = r \cos \theta $. Using the chain rule, we can obtain a system of equations for the evolution of the polar variables as
	\begin{equation}
		\begin{cases}
			\dot r =  \Gamma \left(  (2 \gamma \Gamma - 1)(r^2 -1) \cos\theta - \gamma \Gamma r \sin^2\theta\right),\\
			\dot \theta = -  \frac{\Gamma\sin\theta}{r} (1 - 2 \gamma \Gamma + \gamma \Gamma r \cos\theta)- 2 J \sin \phi,\\
			\dot \phi = - 2 J \cos\phi \cot\theta.
		\end{cases}
	\end{equation}
	Note that the similarity between the equation of motion for $r$ and for the purity, due to $P_t = (1 + r_t^2)/2$ yielding $\dot P_t = r_t \dot r_t$.
	
	Let us now study the steady states of this system of equations. First, note that setting $\phi = \frac{\pi}{2}$, equivalent to $x=0$, trivially gives the steady state of the $\phi$ variable. In the study of nonlinear dynamics \cite{Strogatz2018}, the concept of \textit{nullclines}, defined as the curves where $\dot r = 0, \; \dot \theta = 0, \; \dot \phi=0$, allows for a more intuitive understanding of the behavior of the dynamical system. This is because the steady states of the system live in the intersection of the nullclines. In particular, the nullcline associated to $r$, on which $\dot r=0$, describes the region where purity remains constant in Fig. 1 of the main text, given by
	\begin{equation}
		r^{\ts{n}_r}_{\theta} = \frac{\gamma \Gamma \sin^2 \theta - \sqrt{4 (1 - 2 \gamma \Gamma)^2 \cos^2 \theta + \gamma^2 \Gamma^2 \sin^4 \theta}}{(4 \gamma \Gamma -2) \cos\theta}.
	\end{equation}
	This equation describes the white region of Fig. 1 of the main text with vanishing time-derivative of the purity, where the steady states live. Interestingly, it depends only on the dimensionless product $\gamma \Gamma$ but not on each of the variables independently. In the $\gamma \Gamma \rightarrow \infty$ limit this function converges to 
	\begin{equation}
		r^{\ts{n}_r}_{\theta} = \frac{1 - \cos(2 \theta) - 2 \sqrt{16 \cos^2 \theta + \sin^4 \theta}}{8 \cos\theta}.
	\end{equation}
	The area enclosed by this curve is given by $\frac{1}{2}\int_0^\pi r_\theta^2 \mr d \theta \approx 0.31 \pi$, which means that, in this limit, $31 \%$ of the states in the cut of the Bloch sphere are purifying.
	
	In turn, the nullcline for $\theta$, i.e., the curve that lies in the $(y,z)$ plane of the Bloch sphere in which $\dot \theta =0$, is given by
	\begin{equation}
		r^{\ts{n}_\theta}_{\theta} = \frac{\Gamma (2 \gamma \Gamma - 1) \sin \theta}{ 2 J + \gamma \Gamma^2 \cos \theta \sin \theta}.
	\end{equation}
	This equation does depend on the dimensionless variables $\gamma J$ and $\Gamma/J$ separately and is used to obtain each of the points forming the green line in Fig. 1. }

\section{Comparison of different numerical approaches}\label{app:Comp_dynamics}

We now compare two different numerical resolutions of the SDQ dynamics. The first is to numerically integrate the nonlinear system of differential equations \eqref{eq:EOM_bloch}---we do so by a standard 4th-order Runge-Kutta (RK4) method. The second approach is to get the time evolution of the density matrix from the formal solution of the master equation, i.e., $|\urho_t) = e^{\uL t}|\rho_0)$, where the equation is vectorized to compute the map $e^{\uL t}[\bullet]$ simply through matrix exponentiation. Additionally, this method requires normalization of the state $|\rho_t)= |\urho_t)/\Tr(\urho_t)$, where $\Tr(\urho_t)= (\mbb 1|\urho_t)= \urho_{ee}+\urho_{ff}$ and $\urho_{ee}=\bra{e}\tilde{\rho}_t\ket{e}$. To avoid computing a matrix exponential for each time $t$, we Trotterize the evolution as $e^{\uL t} = \prod_{n=1}^{N_\ts{t}} e^{\uL \Delta_\ts{t}} $, subdividing the evolution in $N_\ts{t}$ steps of length $\Delta_\ts{t} = t_{\rm end}/N_\ts{t}$.

\begin{figure}[h]
	\centering
	\includegraphics[width = .4\linewidth]{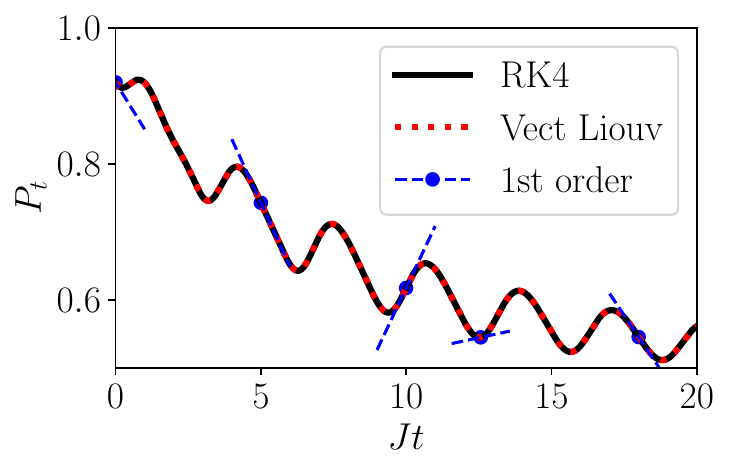}
	\caption{Purity computed from 4th-order Runge Kutta (black line) and using the vectorized Liouvillian (red dotted line). The first-order Taylor approximation, determined from the $\partial_t P$ formula (blue dashed line), matches the two approaches. The parameters for the simulation are $\gamma = 0.5, \; \Gamma = 0.5$ and the initial state has Bloch components $\be r = (0.2, \; 0.8, \; 0.4)$. The absolute value of the difference between the two solutions is of order $10^{-11}$ with a timestep $J \Delta_\ts{t} =0.004$. }
	\label{fig:comparison_purity}
\end{figure}

These two approaches are compared in Fig.~\ref{fig:comparison_purity} to compute the purity, showing a perfect agreement. The first-order evolution of the purity obtained from $\partial_t P$ in eq. (9) of the main text, also agrees with the simulated evolution.

\section{Uhlmann fidelity for a qubit}\label{app:fidelity}
The fidelity between two pure states $\ket{\psi}$ and $\ket{\phi}$ measures how distinguishable the two states are, and is given by $F(\psi, \phi) = |\braket{\psi|\phi}|^2 = \Tr(\ket{\psi}\braket{\psi|\phi}\bra{\phi})$ \cite{nielsen_quantum_2010}. The generalization to any two mixed states $\ha \rho, \; \ha \sigma$ requires the introduction of the \textit{Uhlmann fidelity} \cite{uhlmann_transition_1976,bures_extension_1969} as
\begin{equation}
	F(\ha \rho, \ha \sigma) = \Big(\Tr \sqrt{\sqrt{\ha \rho}\ha \sigma \sqrt{\ha \rho}}\Big)^2.
\end{equation}

This expression is cumbersome to compute due to the matrix square roots, which in particular require dealing with a non-vectorized density matrix. For a two-dimensional Hilbert space, there is a simpler expression for the Uhlmann fidelity given by 
\cite{jozsa_fidelity_1994, hubner_explicit_1992}
\begin{equation}
	F(\ha \rho, \ha \sigma ) = \Tr(\ha \rho \ha \sigma) + 2 \sqrt{\det{\ha \rho}\det{\ha \sigma}}.
\end{equation}
This expression was used in Fig. 3 of the main text to compute the distinguishability between the instantaneous state and the stable steady state.

\section{Comparison of the anti-dephasing Liouvillian with hybrid and tilted Liouvillians}

The Nonlinear Master Equation derived in the main text (4) is not of GKSL form and describes, to the best of our knowledge, an entirely new form of dissipation, what we call anti-dephasing. There are other master equations beyond GKSL form commonly used in the literature; we study their relations in this appendix. 

The \textit{Hybrid Liouvillian} \cite{minganti_hybrid_20} describes the dynamics of a system undergoing continuous monitoring of quantum jumps and post-selection, with a finite-efficiency  $\eta$ detector, and reads, for a single jump operator $\hat L$ with rate $\mu$
\begin{equation}
	\uL_q[\bullet] = -i [\hat H_0, \bullet] + \mu(q \hat L \bullet \hat L\dg - \frac{1}{2}\{\hat L\dg \ha L, \bullet\}),
\end{equation}
which nicely interpolates between Non-Hermitian evolution when the detector is totally efficient $q =1-\eta= 0$ and Lindblad dynamics when the detector is totally inefficient $q=1-\eta = 1$. This hybrid Liouvillian only has a physical interpretation in terms of post-selected trajectories when $q \in [0, 1]$. The dynamics is always trace-decreasing when $q \in [0, 1)$ and trace-preserving for $q=1$.

Another commonly considered master equation beyond GKSL form is the \textit{tilted generator} \cite{garrahan_thermodynamics_2010, carollo_rare_18} or \textit{generalized quantum master equation} \cite{esposito_noneq_09}, whose classical analog is the \textit{Lebowitz-Spohn} operator \cite{Lebowitz1999}. This generator describes the dynamics of a biased ensemble of trajectories and, in its simplest form, reads
\begin{equation}
	\uL_s[\bullet] = - i [\hat H, \bullet] + \mu (e^{-s} \hat L \bullet \hat L\dg - \frac{1}{2}\{\hat L\dg \hat L, \bullet \}).
\end{equation}

This generator describes the dynamics of the biased ensemble of trajectories $\urho_s(t) = \sum_{K=0}^\infty \urho^{(K)}(t) e^{-s K}$, where $\urho^{(K)}$ represents the density matrix of the dynamics with $K$ events after time $t$, i.e. jumps with operator $\ha L$. The variable $s$ represents the conjugate field to $K$. The dynamics generated by this equation is not trace-preserving if $s\neq 0$, if $s<0$ the system is in the \textit{active} phase, in which the trajectories with jumps are favored, and the dynamics is trace-increasing, and if $s>0$ the system is in the \textit{passive} phase with less jumps than usual, and the dynamics is trace-decreasing. 

Both of these generators do not alter the form of the anticommutator term in the GKSL master equation $-\frac{1}{2}\{\ha L\dg \ha L, \bullet\}$. The nonlinear master equation derived in this work, equation (4) of the main text, contains a double anticommutator, which means that this term is positive, instead of negative, as in GKSL. For this reason, the nonlinear master equation does not simply reduce to one of the previously mentioned generators.

There is a particular case in which the connection is closer, if the jump operator is proportional to a projector, i.e., $\ha L \propto \ha \Pi$, where $\ha \Pi^2 = \ha \Pi$, as in the SDQ example studied in the main text (8). In this case, through the mapping $\Gamma - \gamma \Gamma^2 \equiv \tfrac{1}{2}\mu$ and $2 \gamma \Gamma^2 \equiv \mu e^{-s}$ the nonlinear master equation for the qubit can be interpreted as a tilted generator and with $2 \gamma \Gamma^2 \equiv \mu q$ it can be interpreted as a hybrid Liouvillian. However, the two mappings are only valid when the anticommutator term is negative $\Gamma - \gamma \Gamma^2>0$, i.e., $\gamma < \Gamma^{-1}$. Furthermore, the mapping to a hybrid Liouvillian has no physical interpretation when the dynamics is trace increasing, i.e. $\gamma > \frac{1}{2 \Gamma}$.

\section{Standard form of the anti-dephasing master equation}\label{app:standard_form}
We here look for the general structure of master equations describing open systems with balanced gain and loss, specifically, for the generator of the dynamics of an unnormalized density matrix $\urho$
\begin{equation}
\frac{d}{dt}\urho=\uL[\urho],
\end{equation}
where $\uL$ need not be trace-preserving. The normalized density matrix $\hat{\rho}$ evolves according to
\begin{equation}
\frac{d}{dt}\hat{\rho}=\uL[\hat{\rho}]-\tr\big(\uL[\hat{\rho}]\big)\hat{\rho},
\end{equation}
making the equation of motion manifestly nonlinear.
To determine the structure of $\uL$, one can introduce an orthonormal basis of $N$-dimensional operators $\ha F_i=1,\dots,N^2$ such that $\tr(\ha F_i^\dag \ha F_j)=\delta_{ij}$, with $N$ the dimension of the Hilbert space. 
It is convenient to choose $\ha F_{N^2}=\hat{\mathbb{1}}/\sqrt{N}$ so that the $\ha F_i$ are traceless for $i=1,\dots,N^2-1$.
The Liouvillian $\uL$ can be determined analogously to the procedure used to establish the structure of Markovian semigroups and the Lindblad master equation
\cite{Gorini76,BreuerBook}, in our case leading to   
\begin{equation}
\label{genL}
\uL[\ti{\rho}]=-i[\ha H,\ti{\rho}]+\{\ha G,\ti{\rho}\}+\sum_{ij=1}^{N^2}a_{ij}\ha F_i\ti{\rho} \ha F_j^\dag,
\end{equation}
where the Hermitian operator $\hat{H} = i(\ha F-\ha F^\dag)/2$ 
and the operator
\begin{equation}
\ha G=\frac{1}{2N}a_{N^2N^2}\hat{\mathbb{1}}+\frac{1}{2}(\ha F^\dag+\ha F),
\end{equation}
are defined in terms of
\begin{equation}
\ha F=\frac{1}{\sqrt{N}}\sum_{i=1}^{N^2}a_{iN^2}\ha F_i,
\end{equation}
for some positive expansion coefficients $a_{ij}$ ($i,j=1,\dots,N^2$).
The standard Lindblad equation follows from imposing trace preservation in (\ref{genL}) \cite{Gorini76,BreuerBook}, which leads to
\begin{equation}
\ha G=-\frac{1}{2}\sum_{ij=1}^{N^2}a_{ij}\ha F_j^\dag \ha F_i.
\end{equation}
However, by relaxing this condition,
the general structure of $\uL[\hat{\rho}]$ is considered. The rate of norm change, in this case, is
\begin{equation}
\frac{d}{dt}\tr(\srho)=2\tr(\ha G\srho)+\sum_{ij=1}^{N^2}a_{ij}\tr(\ha F_j^\dag \ha F_i\srho).
\end{equation}
It follows that the general structure of the nonlinear master equation for open systems with balanced gain and loss reads
\begin{equation}
\frac{d}{dt}\hat{\rho}-i[\ha H,\hat{\rho}]+\{\ha G,\hat{\rho}\}+\sum_{ij=1}^{N^2}a_{ij}\ha F_i\hat{\rho} \ha F_j^\dag-\Big(2\tr(\ha G\hat{\rho})+\sum_{ij=1}^{N^2}a_{ij}\tr(\ha F_j^\dag \ha F_i\hat{\rho})\Big)\hat{\rho}. 
\end{equation}
This master equation can be brought to a diagonal form by considering the unitary transformation $[\ha u \ha a \ha u^\dag]_{k\ell}=\gamma_k\delta_{k\ell}$ and writing $\ha F_i=\sum_{k=1}^{N^2}u_{ki}\ha A_k$ in terms of the new set of operators $\{\ha A_k\}$, to find
\begin{equation}
\frac{d}{dt}\hat{\rho}=-i[\ha H,\hat{\rho}]+\{\ha G,\hat{\rho}\}+\sum_{ij=1}^{N^2}\gamma_k \ha A_k\hat{\rho} \ha A_k^\dag -\Big(2\tr[\ha G\hat{\rho}]+\sum_{k=1}^{N^2}\gamma_k\tr[\ha A_k^\dag \ha A_k\hat{\rho}]\Big)\hat{\rho},
\end{equation}
with 
\begin{equation}
\ha G=\frac{1}{2N}a_{N^2N^2}\hat{\mathbb{1}}+\frac{1}{2\sqrt{N}}\gamma_{N^2}(\ha A_{N^2}+\ha A_{N^2}^\dag).
\end{equation}

\end{document}